\documentclass[prl,aps,preprint,showpacs,preprintnumbers,amsmath,amssymb,secnumroman,eqsecnum,eqappnum]{revtex4}

\usepackage{graphicx}
\usepackage{dcolumn}
\usepackage{bm}

\newcommand{\beq}{\begin{equation}}
\newcommand{\eeq}{\end{equation}}
\newcommand{\beqa}{\begin{eqnarray}}
\newcommand{\eeqa}{\end{eqnarray}}
\newcommand{\vc}[1]{\mbox{\boldmath $#1$}}

\begin{document}


\title{Comment on:"Scallop Theorem and Swimming at the Mesoscale" by M. Hubert, O. Trosman, Y. Collard, A. Sukhov, J. Harting, N. Vandewalle, and A.-S. Smith, Phys. Rev. Letters \vc{126}, 224501 (2021)}

\author{B. U. Felderhof}
\email{ufelder@physik.rwth-aachen.de}

\affiliation{Institut f\"ur Theorie der Statistischen Physik\\ RWTH Aachen University\\
Templergraben 55\\52056 Aachen\\ Germany\\
}%



\date{\today}

\maketitle
The authors did not compare with the exact expression for the swimming velocity which we derived earlier $[25]$.   We consider the two-sphere swimmer with prescribed stroke $x_2(t)-x_1(t)=L+d\sin(\omega t)$. In Ref. 25 we derived expressions for the corresponding center velocity $U(t)=dC/dt,\;C(t)=(x_1(t)+x_2(t))/2$, and the mean swim velocity $\overline{U}_{sw}$, i.e. the average of $U(t)$ over a period $T=2\pi/\omega$. The expressions can be evaluated numerically to any desired accuracy.\\

The expressions follow from the formal solution of the swim equation [1]
\begin{equation}
\label{A.1}\frac{d}{dt}(MU)+ZU=\mathcal{I},
\end{equation}
where $M$ is the mass, $Z$ is the friction coefficient, and $\mathcal{I}$ is the impetus. In the present model the mass $M$ is time-independent. The friction coefficient and the impetus are time-dependent and can be calculated from the stroke.  The zeroth harmonic of $U(t)$ is the result of rectification, and is identified as the mean swim velocity $\overline{U}_{sw}$.

One gains qualitative understanding by solving Eq. (0.1) in first harmonic approximation, taking account of only zeroth and first harmonics. In first harmonic approximation the mean swim velocity is denoted as $\overline{U}^{(1)}_{sw}$. This takes the form
 \begin{equation}
\label{A.2}\overline{U}^{(1)}_{sw}=\omega^2\eta\bigg[\frac{\eta^2}{uM+vM_d}+\frac{M^2\omega^2}{a_1a_2(wM+zM_d)}\bigg]^{-1},
\end{equation}
where $M_d=m_1-m_2$ is the mass difference. The coefficients $u,v,w,z$ are complicated dimensionless functions of the four lengths $a_1,a_2,L,d$.

For small $d$ and large $L$ the coefficients are given approximately by
\begin{eqnarray}
\label{A.3}u_2&=&-\frac{a_1a_2(a_1-a_2)}{8\pi(a_1+a_2)^3}\frac{d^2}{L^2},\qquad v_2=\frac{a_1a_2}{8\pi(a_1+a_2)^2}\frac{d^2}{L^2},\nonumber\\
w_2&=&-\frac{9\pi}{2}\frac{a_1-a_2}{a_1+a_2}\frac{d^2}{L^2},\qquad z_2=\frac{9\pi}{2}\frac{d^2}{L^2}.
\end{eqnarray}
With these values $\overline{U}^{(1)}_{sw2}$ is identical with $\overline{U}^{S}$ in Eq. (3).

For large viscosity $\eta$ the first term in square brackets in Eq. (0.2) dominates. This leads to
 \begin{equation}
\label{A.4}\overline{U}^{(1)}_{sw2}\approx\frac{1}{3}\frac{a_1^2a_2^2}{(a_1+a_2)^3}|\rho_1a_1^2-\rho_2a_2^2|\frac{\omega^2d^2}{\eta L^2},\qquad(\mathrm{large}\;\eta,\;\mathrm{small}\; d/L ).
\end{equation}
For $\rho_1=\rho_2$ this agrees with Eq. (5). For small viscosity the second term in square brackets dominates. This leads to
 \begin{equation}
\label{A.5}\overline{U}^{(1)}_{sw2}\approx\frac{3}{16\pi^2}\frac{a_1^2a_2^2}{(a_1+a_2)^3}\frac{|\rho_1a_1^2-\rho_2a_2^2|}{(\rho_1a_1^3+\rho_2a_2^3)^2}
\frac{\eta d^2}{L^2},\qquad(\mathrm{small}\;\eta,\;\mathrm{small}\; d/L ).
\end{equation}
The second term also dominates for large $\rho_1,\rho_2$, leading to the same expression. We disagree with the result (4) for this case. The mean swimming velocity tends to zero as the mass densities increase, rather than to infinity.

For intermediate values of the parameters the complete expression (0.2) must be used. We measure viscosity in terms of the number $R=a_1^2\omega\rho_1/\eta$. In Fig. 1 we plot $10^3\overline{U}^{(1)}_{sw}/(\omega a_1)$ as a function of $\log_{10}R$ for the case $a_1=1,\;a_2=0.5,\;L=3,\;d=1,\;\omega=1,\;\rho_1=\rho_2=1$, and compare with the approximation $\overline{U}^{(1)}_{sw2}=\overline{U}^{S}$, as well as with the exact expression Eq. (3.10) of Ref. 25 for $\overline{U}_{sw}$. The plot shows that even for this relatively large value of $d/L$ the first harmonic approximation $\overline{U}^{(1)}_{sw}$ is quite accurate. For smaller values of $d/L$ the difference between $\overline{U}^{S}$ and $\overline{U}_{sw}$ gets smaller and the first harmonic approximation $\overline{U}^{(1)}_{sw}$ gets even closer to the exact value $\overline{U}_{sw}$.\\\\

\newpage

\section*{Figure caption}

\subsection*{Fig. 1}
Plot of the mean swim velocity $\overline{U}_{sw}$ of the two-sphere model with parameters $a_1=1,a_2=0.5,L=3,d=1$ as a function of $R=a_1^2\omega\rho_1/\eta$ (solid curve). We compare with the mean swim velocity $\overline{U}^{(1)}_{sw}$ calculated in first harmonic approximation (short dashes), and with the approximation $\overline{U}^{(1)}_{sw2}=\overline{U}^S$ (long dashes).

\newpage
\setlength{\unitlength}{1cm}
\begin{figure}
 \includegraphics{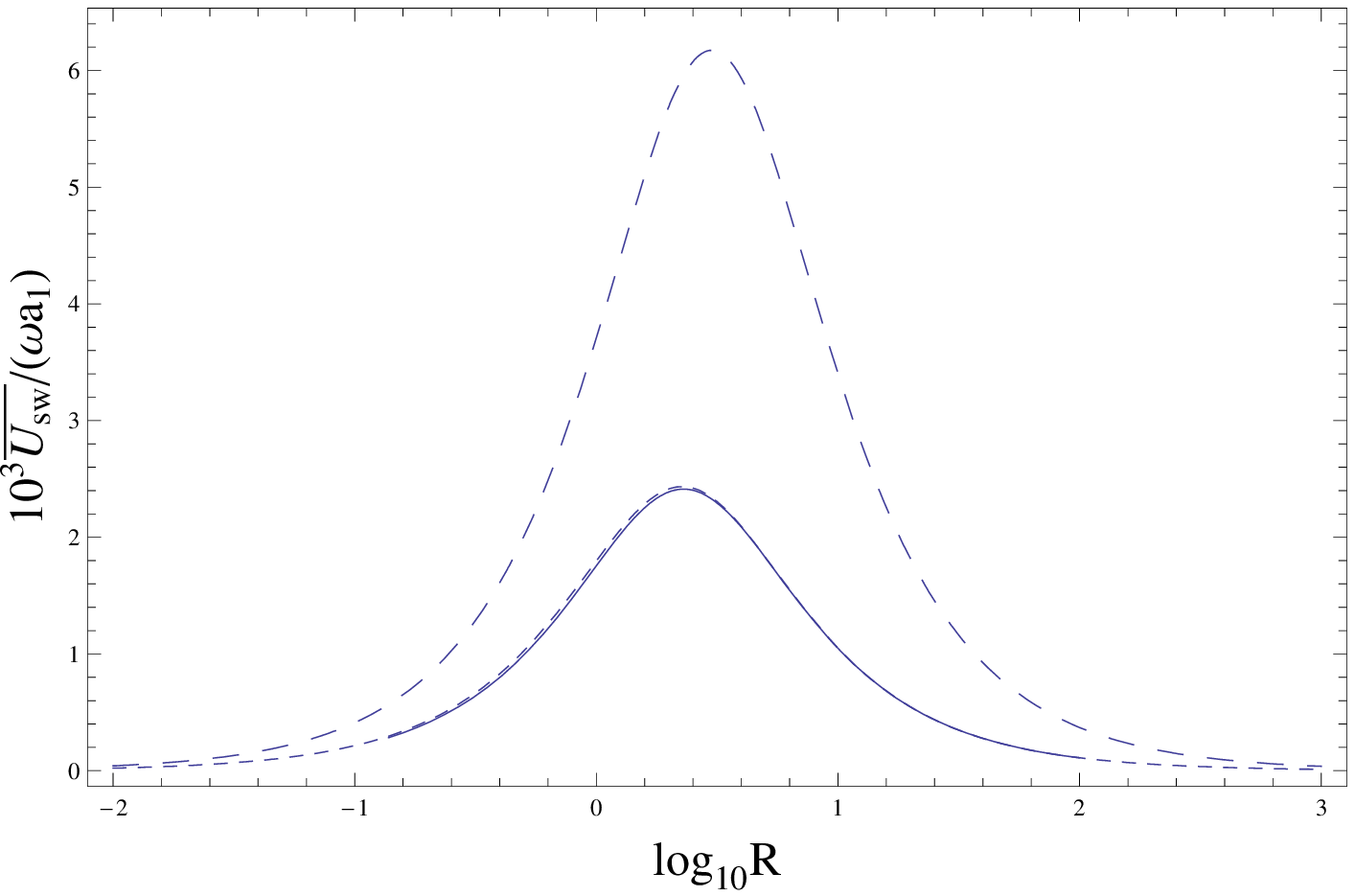}
   \put(-9.1,3.1){}
\put(-1.2,-.2){}
  \caption{}
\end{figure}
\end{document}